\documentclass[printer]{aa}

\usepackage[varg]{txfonts}
\usepackage{amsmath}
\usepackage{graphicx}
\usepackage{comment}
\usepackage{color} 
\usepackage{natbib}
\newcommand{\fig}[1]{Fig.~\ref{#1}}
\newcommand{\tbl}[1]{Table~\ref{#1}}
\newcommand{\sect}[1]{Sect.~\ref{#1}}
\newcommand{\p}{Paper I}

\begin{document}
\title{How flux feeding causes eruptions of solar magnetic flux ropes with the hyperbolic flux tube configuration?}
\author{
Quanhao Zhang\inst{\ref{inst1},\ref{inst2},\ref{inst3}}\and 
Rui Liu\inst{\ref{inst1},\ref{inst2},\ref{inst4}}\and 
Yuming Wang\inst{\ref{inst1},\ref{inst2},\ref{inst3}}\and 
Zhenjun Zhou\inst{\ref{inst5}}\and 
Bin Zhuang\inst{\ref{inst6}}\and 
Xiaolei Li\inst{\ref{inst1},\ref{inst3}}}
\institute{CAS Key Laboratory of Geospace Environment, School of Earth and Space Sciences, University of Science and Technology of China, Hefei 230026, China\\ \email{zhangqh@ustc.edu.cn}\label{inst1}
\and
CAS Center for Excellence in Comparative Planetology, University of Science and Technology of China, Hefei 230026, China\label{inst2}
\and
Mengcheng National Geophysical Observatory, School of Earth and Space Sciences, University of Science and Technology of China, Hefei 230026, China\label{inst3}
\and
Collaborative Innovation Center of Astronautical Science and Technology, Hefei, Anhui 230026, China\label{inst4}
\and
School of Atmospheric Sciences, Sun Yat-sen University, Zhuhai, Guangdong, 519000, China\label{inst5}
\and
Institute for the Study of Earth, Ocean, and Space, University of New Hampshire, Durham, NH 03824, USA\label{inst6}
}
\abstract{Coronal magnetic flux ropes are generally considered to be the core structure of large-scale solar eruptions. Recent observations found that solar eruptions could be initiated by a sequence of ``flux feeding," during which chromospheric fibrils rise upward from below, and merge with a pre-existing prominence. Further theoretical study has confirmed that the flux feeding mechanism is efficient in causing the eruption of flux ropes that are wrapped by bald patch separatrix surfaces. But it is unclear how flux feeding influences coronal flux ropes that are wrapped by hyperbolic flux tubes (HFT), and whether it is able to cause the flux-rope eruption. In this paper, we use a 2.5-dimensional magnetohydrodynamic model to simulate the flux feeding processes in HFT configurations. It is found that flux feeding injects axial magnetic flux into the flux rope, whereas the poloidal flux of the rope is reduced after flux feeding. Flux feeding is able to cause the flux rope to erupt, provided that the injected axial flux is large enough so that the critical axial flux of the rope is reached. Otherwise, the flux rope system evolves to a stable equilibrium state after flux feeding, which might be even farther away from the onset of the eruption, indicating that flux feeding could stabilize the rope system with the HFT configuration in this circumstance.}
\titlerunning{Coronal eruptions caused by flux feeding in HFT configurations}
\maketitle

\section{Introduction}
\label{sec:introduction}
Large-scale solar eruptions include prominence/filament eruptions, flares, and coronal mass ejections (CMEs) \citep{Benz2008a,Chen2011a,Parenti2014a,Liu2020}. They are capable of inflicting huge impacts on the solar-terrestrial system \citep{svestka2001a,Cheng2014,Shen2014,Lugaz2017,Gopalswamy2018a}. It is widely accepted that different kinds of large-scale solar eruptions are close related to each other: they are essentially different manifestations of the same eruptive process of a coronal magnetic flux rope system \citep{Zhang2001,Vrvsnak2005a,vanDriel2015a,Jiang2018,Liu2018a,Yan2020}. Therefore, it is of great significance to investigate how the eruption of coronal magnetic flux ropes is initiated. According to the magnetic topology, coronal flux ropes are classified into two types of configurations: if the flux rope sticks to the photosphere, with a bald patch separatrix surface \citep[BPSS,][]{Titov1993a,Titov1999a,Gibson2006a} wrapping the flux rope, this is usually called the BPS configuration \citep{Filippov2013}; for the flux rope system in which the rope is suspended in the corona and wrapped around by a hyperbolic flux tube (HFT), it is called the HFT configuration \citep{Titov2003,Aulanier2005,Chintzoglou2017}.
\par
Many theoretical analyses have been carried out to investigate the eruptive mechanism of coronal magnetic flux ropes. Apart from the well-known magnetohydrodynamic (MHD) instabilities \citep{Romano2003,Torok2003a,Kliem2006a,Fan2007a,Aulanier2010a,Guo2010,Savcheva2012b}, it was also suggested by many previous studies that catastrophes could be responsible for solar eruptions \citep[e.g.,][]{Forbes1991a,Isenberg1993a,Lin2001a,Chen2007a,Demoulin2010a,Longcope2014a,Kliem2014}. For example, if either the axial flux or the poloidal flux of a flux rope exceeds the corresponding critical value, a catastrophic loss of equilibrium will occur, resulting in the eruption of the rope \citep{Li2002a,Bobra2008,Su2011a,Zhang2016a,Zhuang2018}. Alternatively, magnetic reconnection may play a dominant role in flux rope eruptions, such as break-out model \citep{Antiochos1999a,Sterling2004}, tether-cutting model \citep{Moore2001a,Inoue2015}, and flux emergence model \citep{Chen2000a,Archontis2008b}. Recently, \cite{Zhang2014} observed that chromospheric fibrils rise from below a quiescent prominence and merge with this prominence. This phenomenon is called ``flux feeding" process. As observed by \cite{Zhang2014}, the flux feeding processes occur 3 times, followed by the eruption of the quiescent prominence. This implies that coronal eruptions could be initiated by flux feeding.
\par
Since there lacks accurate measurement of the local conditions in the corona, the physical essence of flux feeding mechanism is unclear based on observational results alone. To solve this, \cite{Zhang2020} (hereafter \p) carried out numerical simulations to investigate the scenario of the flux feeding mechanism in BPS configurations. In their simulations, the rising chromospheric fibril is represented by a small flux rope emerging from below a pre-existing coronal magnetic flux rope with a BPS configuration. It was found that flux feeding processes only inject axial magnetic flux into the pre-existing flux rope, whose poloidal flux, however, remains unchanged after flux feeding. If the amount of the injected axial flux is large enough so that the total axial flux of the flux rope exceeds its critical value, the eruption of the rope is initiated by the upward catastrophe associated with the critical axial flux; otherwise, the flux rope remains in the stable BPS configuration after flux feeding. Different from the BPS configurations, HFT configurations are usually considered as the pre-eruptive states of coronal eruptions \citep{Galsgaard2003,Aulanier2010a,Filippov2013}, indicating that HFT configurations could be metastable or even unstable. Here come the questions: how does flux feeding influence the flux rope systems with an HFT configuration? Is flux feeding still able to cause eruptions? If flux feeding mechanism is still efficient, what are the similarities and differences between the BPS and the HFT cases? To resolve these questions, theoretical analyses about flux feeding in HFT configurations are needed, which will further shed light on the physical nature of the flux feeding mechanism.
\par
In this paper, we carry out 2.5-dimensional numerical simulations to investigate the influence of flux feeding on flux rope system with an HFT configuration, especially how flux feeding causes the rope's eruption. The rest of the paper is arranged as follows: the simulating procedures are introduced in \sect{sec:method}; the simulation results of a typical flux feeding process and the flux rope eruption caused by this process are presented in \sect{sec:result}; the influence of flux feeding on the rope system and the physical scenario of flux feeding mechanism in HFT configurations are analyzed in \sect{sec:analysis}. Finally, discussion and conclusion are given in \sect{sec:dc}.

\section{Simulating procedures}
\label{sec:method}
\begin{figure*}
\includegraphics[width=\hsize]{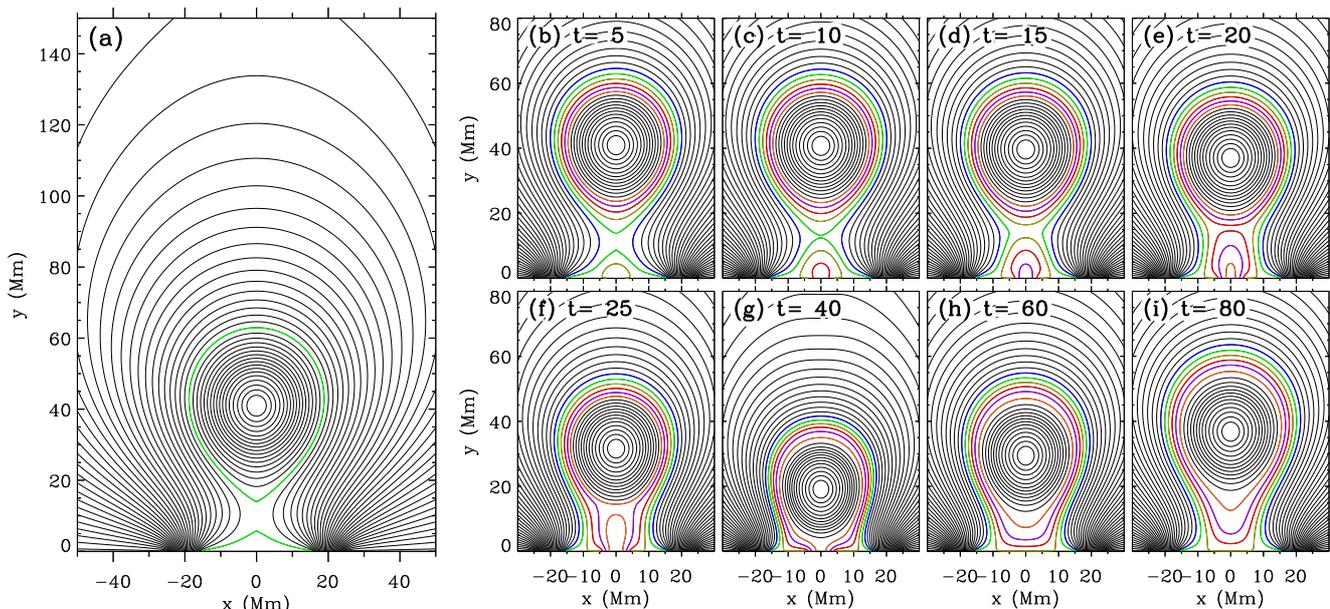}
\caption{Evolution of the magnetic configuration during a flux feeding process with $C_E=2.23$. Panel (a) shows the initial flux rope system with an HFT configuration; the green curves mark the boundary of the major flux rope and that of the arcades below the rope. The times marked in panels (b)-(i) are in the unit of $\tau_A$. Different colors depict the magnetic field lines corresponding to different values of $\psi$.}\label{fig:feeding}
\end{figure*}

For the 2.5-dimensional cases in our simulations, all the magnitudes satisfy $\partial/\partial z=0$, thus the magnetic field can be denoted as
\begin{align}
\textbf{B}=\triangledown\psi\times\hat{\textbf{\emph{z}}}+B_z\hat{\textbf{\emph{z}}}.\label{equ:mf}
\end{align}
Here $B_z$ is the component of the magnetic field in $z-$dirention; $\psi$ is the magnetic flux function, and the isolines of $\psi$ correspond to the magnetic field lines projected in $x-y$ plane. In this paper, the multi-step implicit scheme \citep{Hu1989a} is used to simulate the evolution of coronal flux rope systems. The initial state in our simulation is obtained by numerical procedures: first we insert a flux rope into a potential background field from the lower base, and then by adjusting the properties of this rope, a flux rope system with an HFT configuration is eventually obtained. The basic equations and the detailed simulating procedures are introduced in Appendix \ref{ape:equations}. The magnetic configuration of the initial state is shown in \fig{fig:feeding}(a): a flux rope is suspended above a bundle of closed arcades, resulting in a X-type magnetic structure below the rope. The green curves in \fig{fig:feeding}(a) mark the boundary of the flux rope and that of the arcades below the rope. The magnetic flux of these arcades per unit length along the $z-$direction is $\Phi_a=0.069\times10^{10}~\mathrm{Mx}~\mathrm{cm}^{-1}$. The background field is a partially open bipolar field, with a negative and a positive surface magnetic charge located at the lower photosphere within $-25~\mathrm{Mm}<x<-15~\mathrm{Mm}$ and $15~\mathrm{Mm}<x<25~\mathrm{Mm}$, respectively. $B_z$ is zero in the background field.
\par
Similar as that in \p, the emerging fibril in the scenario of flux feeding is represented by a small flux rope, and the pre-existing large flux rope is called ``major rope" for simplicity in the rest of this paper. The emergence of the small rope is achieved by similar simulating procedures as those in \p: the small rope emerges from the central region right below the major rope at a constant speed; the emergence begins at $t=0$ and ends at $t=\tau_E=60 \tau_A$, where $\tau_A=17.4$ s is the characteristic Alfv\'{e}n transit time. Thus the emerged part of the small rope at the base is located within $-x_E\leqslant x\leqslant x_E$, where $x_E=(a^2-h_E^2)^{1/2}$, $h_E=a(2t/\tau_E-1)$, $0\leqslant t\leqslant \tau_E$, and $a$ is the radius of the small rope. Based on this, by adjusting $\psi$, $B_z$, the velocities $v_{x,y,z}$, the temperature $T$, and the density $\rho$ at the base ($y=0, -x_E\leqslant x\leqslant x_E$), the emergence of the small flux rope is achieved by:
\begin{align}
&\psi(t,x,y=0)=\psi_i(x,y=0)+\psi_E(t,x),\\
&\psi_E(t,x)=\frac{C_E}{2}\mathrm{ln}\left(\frac{2a^2}{a^2+x^2+h_E^2}\right)\label{equ:psi},\\ 
&B_z(t,x,y=0)=C_Ea(a^2+x^2+h_E^2)^{-1}\label{equ:bz},\\
&v_y(t,x,y=0)=v_E=2a/\tau_E,~v_x(t,x,y=0)=v_z(t,x,y=0)=0,\\
&T(t,x,y=0)=2\times10^5\mathrm{~K},~\rho(t,x,y=0)=1.67\times10^{-12}\mathrm{~kg~m^{-3}}.
\end{align}
Here $v_E=2a/\tau_E$ is the emerging velocity; $\psi_i$ is the magnetic flux function of the initial state, and it is determined by numerical means (please see Eq. \ref{equ:fluxb} in Appendix \ref{ape:equations}). It is noteworthy that $\psi$ at the lower base is fixed at $\psi_i$ except during the emergence of the small rope, so that the normal component of the photospheric magnetic field is unchanged, implying that the lower base corresponds to the photosphere \citep{Lin2003a}. The parameter $C_E$ determines the magnetic field strength of the small rope; its dimensionless values given in this paper are in the unit of $0.373\times10^10\mathrm{~Mx~cm^{-1}}$. In both the small and the major ropes, $B_z$ is positive and the component of the magnetic field in $x$-$y$ plane is counterclockwise. Anomalous resistivity is used in our simulations, so as to restrict magnetic reconnection within the regions of current sheets:
\begin{align}
\eta=
\begin{cases}
0,& ~j\leq j_c\\
\eta_m\mu_0v_0L_0(\frac{j}{j_c}-1)^2.& ~j> j_c \label{equ:res}
\end{cases}
\end{align}
Here $\eta_m=9.95\times10^{-2}$, $L_0=10^7$ m, and $v_0=128.57$ km s$^{-1}$; $\mu_0$ is the vacuum magnetic permeability; the critical current density is $j_c=4.45\times10^{-4}$ A m$^{-2}$.

\section{Simulation results}
\label{sec:result}

\subsection{Flux feeding process}
\label{sec:feeding}
With the procedures introduced above, we are able to simulate flux feeding processes in HFT configurations. The simulation results of a typical flux feeding process  with $C_E=2.23$ are illustrated in \fig{fig:feeding}(b)-\ref{fig:feeding}(i); different colors are used to depict the magnetic fields lines corresponding to different values of the flux function $\psi$, so as to clearly demonstrate the interaction between the major rope and emerging small rope in detail.
\par
\begin{figure*}
\includegraphics[width=\hsize]{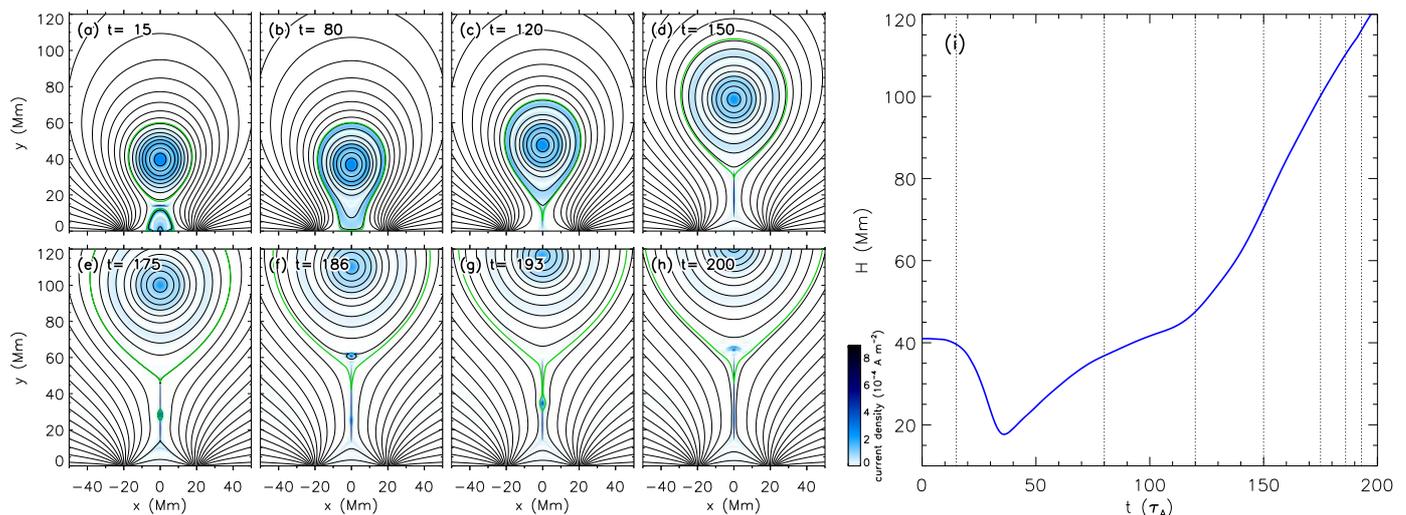}
\caption{Eruptive process of the resultant flux rope after the flux feeding process with $C_E=2.23$. Panels (a)-(h) illustrate the magnetic configurations at different times, and the blue color depicts the distribution of the current density. Panel (i) plots the evolution of the height of the rope axis, and the times of Panels (a)-(h) are marked by the vertical dotted lines in panel (i). The green curves in panels (a)-(h) mark the outer boundaries of the flux ropes and the plasmoids.}\label{fig:erupt}
\end{figure*}
At the beginning of the flux feeding process, the small rope starts to emerge from the lower boundary, as marked by the brown curve in \fig{fig:feeding}(b). The arcades below the major rope in the initial state are then pushed upward by the emerging small rope, so that the top boundary of these arcades reaches the lower boundary of the major rope (as shown by the green curves in \fig{fig:feeding}(c)), and then these arcades reconnect with the magnetic fields of the major rope (\fig{fig:feeding}(d)). A comparison between \fig{fig:feeding}(d) and \fig{fig:feeding}(b) shows that the outer section of the major rope is peeled off by the reconnection. After all the arcades have reconnected with the major rope, the small rope itself also reaches the major rope. During the first half ($0\sim\tau_E/2=30\tau_A$) of the flux feeding process, what emerges from the lower base is the top half of the small rope, in which the direction of $B_x$ is opposite to that in the bottom half of the major rope. Thus a horizontal current sheet forms at the interface between the two ropes. This current sheet can be clearly recognized in \fig{fig:erupt}(a), in which the blue color depicts the distribution of the current density. As a result of the reconnection within this current sheet, the two ropes gradually merge together, as shown by, for example, the magnetic field lines plotted by the red and the pink colors in \fig{fig:feeding}(e)-\ref{fig:feeding}(f). The magnetic cancelation occurring in this current sheet should decrease the local magnetic pressure below the major rope, so that the major rope gradually descends during the early period of the flux feeding process (see \fig{fig:feeding}(f) and \ref{fig:feeding}(g), also see the height profile of the major rope axis plotted in \fig{fig:erupt}(i)). During the second half ($30\tau_A\sim60\tau_A$) of the flux feeding process, however, it is the bottom half of the small rope that emerges from the lower base; the direction of the magnetic field lines in the small rope is now the same as that in the bottom half of the major rope; there should be no reconnection any more. Therefore, after $t=30\tau_A$, magnetic flux is injected from the lower base, so that the local magnetic pressure accumulates below the major rope, which causes the major rope to gradually rise, as shown by \fig{fig:feeding}(g)-\ref{fig:feeding}(h).
\par
Eventually, the flux feeding process ends at $t=60\tau_A$. The configuration of the resultant flux rope after flux feeding is illustrated in \fig{fig:feeding}(i): the rope sticks to the photosphere, without any arcade below the rope. 
\par

\subsection{Eruption caused by flux feeding}
\label{sec:erupt}
Further evolution of the resultant major rope after flux feeding indicates that the major rope is caused to erupt by this flux feeding process with $C_E=2.23$. As demonstrated by \fig{fig:erupt}, the major rope keeps rising after flux feeding. Due to the line-tying effect from the photosphere, the rope does not separate from the lower base immediately, but its lower boundary sticks to the photosphere for a certain period (\fig{fig:erupt}(b)). As the rope rises, the lower section of the flux rope and the ambient background field are stretched. Eventually, as shown in \fig{fig:erupt}(c)-\ref{fig:erupt}(d), the rope is fully detached from the lower base, and magnetic reconnection occurs within the vertical current sheet formed below the rope. The reconnection results in closed arcades below the rope, which can be clearly recongized in the lower section of \fig{fig:erupt}(d). Moreover, as shown in \fig{fig:erupt}(e), a plasmoid appears within the current sheet below the rope, and then rises and merges with the major rope (\fig{fig:erupt}(f)), which is similar as the observations and simulations in \cite{Gou2019}.  This process is followed by another similar one, as shown in \fig{fig:erupt}(g)-\ref{fig:erupt}(h). These plasmoids should result from the tearing mode instability within the current sheet, which have been investigated in detail in many previous studies \cite[e.g.,][]{Barta2008,Shen2011,Ni2012,Huang2017}. 
\par
The height of the major rope axis versus time is plotted in \fig{fig:erupt}(i). As explicated in \sect{sec:feeding}, due to the magnetic cancellation during the early period of the flux feeding process, the major flux rope gradually descends, which is accompanied by the contraction of the background arcades above the major rope (see \fig{fig:feeding}(f) and \ref{fig:feeding}(g)). This kind of downward movement before the onset of the eruption is often considered as a signature of ``coronal implosion" \citep{Hudson2000}, which has been reported in many previous observational studies \cite[e.g.,][]{Ji2004,Veronig2006,Joshi2009,Liu2009,Li2017}. For example, \cite{Liu2009} observed the large-scale contraction of the overlying coronal loops before the onset of the flare, which was attributed to the prolonged preheating phase dominated by coronal thermal emissions. Our simulation results suggest another scenario of coronal implosions: internal magnetic cancellation during the early period of the flux feeding process in HFT configurations decreases the local magnetic pressure below the major flux rope, resulting in the downward movement of the rope before the onset of its eruption. For comparison, the height profile of the flux rope eruption caused by flux feeding in BPS configurations is monotonic, without any downward movement before the eruption (see \p).
\par
\begin{figure*}
\includegraphics[width=\hsize]{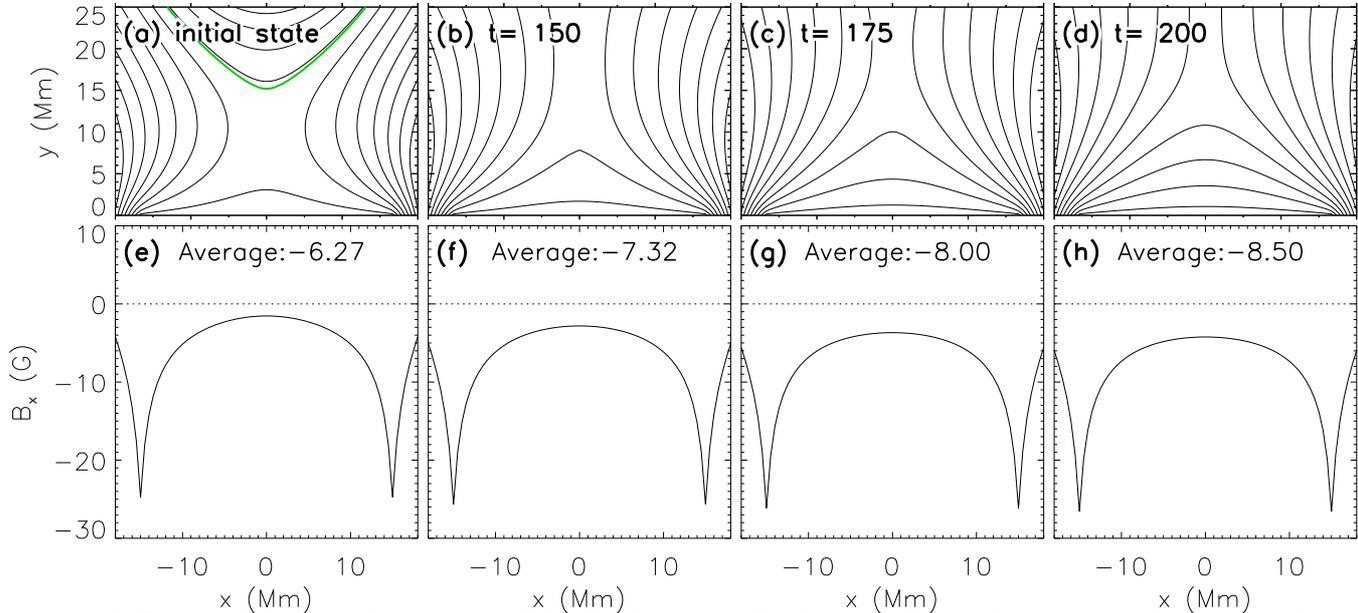}
\caption{Variation of the photospheric magnetic field after the eruption. Panels (a)-(d) illustrate the magnetic configurations at the central region above the PIL, and panels (e)-(h) plot the corresponding horizontal component of the photospheric magnetic fields, $B_x$. The times marked in panels (b)-(d) are in the unit of $\tau_A$.}\label{fig:bx}
\end{figure*}

It has been suggested in many previous studies that solar eruptions not only could be initiated by the gradual variations of photospheric magnetic fields, but also might in turn cause rapid magnetic field changes in the photosphere \citep{Wang1994,Liu2005,Wang2015,Toriumi2019}. The issue about this kind of photospheric response to the coronal eruptions is graphically called ``tail wags the dog" problem \citep{Aulanier2016}. This phenomenon has been observed in many previous studies \cite[e.g.,][]{Wang2012,Liu2012c,Petrie2012,Sun2017,CastellanosDuran2018}. For example, by measuring the various magnetic parameters of a compact region at the central polarity inversion line (PIL) of an active region, \cite{Sun2017} found that the horizontal component of the photospheric magnetic field in this compact region obviously increase after the flare. In our simulation results, \fig{fig:bx} shows the photospheric distribution of the horizontal magnetic field, $B_x$, at the central region right below the major rope (we note that $B_z$ is zero in the background field). In the initial state, as shown in \fig{fig:bx}(e), $B_x$ at the photosphere is negative, consistent with the arcades below the rope in the initial state. During the eruptive process of the resultant major rope after flux feeding, closed arcades forms right below the rising rope, which also results in the negative photospheric $B_x$, as shown in \fig{fig:bx}(f). Moreover, as the rope rises, more and more field lines are closed by reconnection and pile up above the PIL. As a result of the reconnection-driven contraction of the flare loops, the horizontal photospheric magnetic field $B_x$ increases accordingly (see \fig{fig:bx}(f)-\ref{fig:bx}(h)), which is consistent with the simulation results in \cite{Barczynski2019}. Comparing the photospheric $B_x$ at $t=200\tau_A$ with that in the initial state, the average strength of the photospheric $B_x$ after the eruption increases from 6.27 G to 8.50 G.  
Moreover, as shown in \fig{fig:bx}, the magnetic field strength is of the order of 10 G, indicating that coronal eruptions originating in quiescent regions could also result in the phenomenon of ``tail wags the dog", that is coronal eruptions originating in quiescent regions could also cause variations of the photospheric magnetic fields.

\section{Initiation of Eruptions}
\label{sec:analysis}
In order to investigate how the eruption is initiated, we first investigate the influence of flux feeding on the major flux rope. The magnetic properties of a flux rope could be characterized by its axial magnetic flux, $\Phi_z$, and its poloidal magnetic flux per unit length along the $z-$direction, $\Phi_p$. For the initial major rope shown in \fig{fig:feeding}(a), the axial flux $\Phi_{z0}=4.366\times10^{19}~\mathrm{Mx}$, and the poloidal flux $\Phi_{p0}=1.186\times10^{10}~\mathrm{Mx}~\mathrm{cm}^{-1}$. Here we select the flux rope at $t=80\tau_A$ (\fig{fig:feeding}(i)) as the resultant rope after flux feeding, whose axial flux $\Phi_{z}=7.115\times10^{19}~\mathrm{Mx}$, and poloidal flux $\Phi_{p}=1.116\times10^{10}~\mathrm{Mx}~\mathrm{cm}^{-1}$. The increased axial flux is injected by the flux feeding process, which is similar as the BPS cases investigated in \p. Different from the result in BPS cases, however, the poloidal flux is reduced by $0.070\times10^{10}~\mathrm{Mx}~\mathrm{cm}^{-1}$ after flux feeding. 
\par
Since the distribution of the coronal and the photospheric magnetic fields play a dominant role in triggering coronal flux rope eruptions \citep{Yeates2014,Yang2018,Thalmann2019,Xing2020}, the influence of flux feeding processes on the major rope system should be sensitive to the scale of the magnetic field strength in the small rope. The magnetic parameters of the resultant flux ropes after the flux feeding processes with different $C_E$ are tabulated in \tbl{tbl:para}. Larger $C_E$ implies stronger magnetic field strength in the small emerging rope, so that more axial flux is injected into the major rope. The deduced poloidal fluxes in different cases, however, are very close to each other: $\Delta\Phi_p=0.069\sim0.070\times10^{10}~\mathrm{Mx}~\mathrm{cm}^{-1}$, which is almost the same as the magnetic flux of the arcades below the rope in the initial state, $\Phi_a$. This indicates that the reduced poloidal flux should be caused by the interaction between the major rope and the arcades below the rope with the HFT configuration, which has been introduced in \sect{sec:feeding}: as pushed by the rising small flux rope, these arcades reconnect with the major rope, so that the outer section of the major rope is peeled off. 
\begin{table}
\caption{The parameters of the resultant flux ropes after the flux feeding processes with different $C_E$.}
\label{tbl:para}
\centering
\begin{tabular}{c c c c}
\hline
$C_E$ & $\Phi_z$ ($10^{19}$ Mx) & $\Phi_p$ ( $10^{10}~\mathrm{Mx}~\mathrm{cm}^{-1}$) & Erupt or not \\
\hline
Initial  & 4.366  & 1.186  & N\\
1.80     & 5.431  & 1.116  & N\\
1.90     & 5.684  & 1.116  & N\\
2.00     & 5.990  & 1.117  & N\\
2.10     & 6.426  & 1.117  & N\\
2.20     & 6.929  & 1.117  & N\\
2.22     & 7.051  & 1.117  & N\\
2.23     & 7.115  & 1.116  & Y\\
2.30     & 7.562  & 1.116  & Y\\
2.40     & 8.281  & 1.116  & Y\\
\hline
\end{tabular}
\tablefoot{$\Phi_z$ and $\Phi_p$ are the axial flux and the poloidal magnetic flux per unit length along the $z-$direction of the major rope, respectively.}
\end{table}
\par
\begin{figure*}
\includegraphics[width=\hsize]{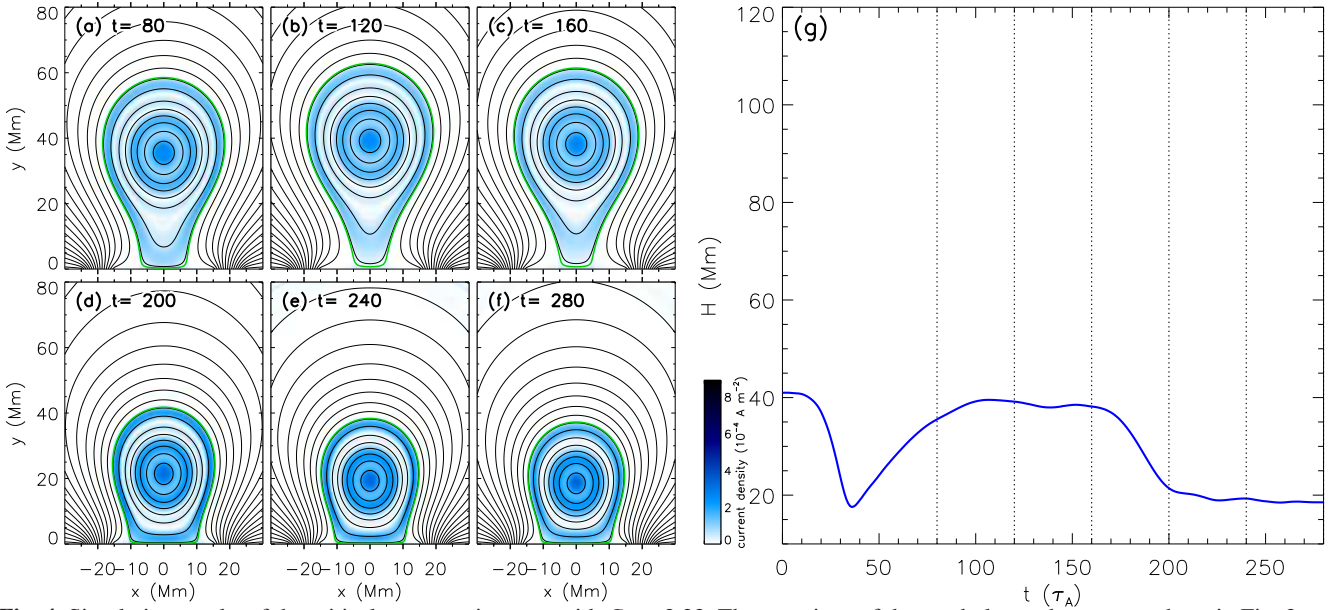}
\caption{Simulation results of the critical non-eruptive case with $C_E=2.22$. The meanings of the symbols are the same as those in \fig{fig:erupt}.}\label{fig:down}
\end{figure*}

In the cases with differrent $C_E$, the further evolution of the resultant rope also varies, as tabulated in the last column in \tbl{tbl:para}. The eruptive and the non-eruptive cases are well separated: the resultant rope erupts if $C_E$ is no smaller than 2.23, and the axial fluxes of the resultant ropes in all the eruptive cases are larger than those in the non-eruptive cases. This indicates that there should exist a critical axial magnetic flux of the resultant rope, and the onset of its eruption should be initiated by the upward catastrophe associated with this critical axial flux \citep[e.g.,][]{Su2011a,Zhang2016a}. The value of the critical axial flux is of the order of $7.1\times10^{19}~\mathrm{Mx}$. The simulation results of the critical non-eruptive case with $C_E=2.22$ are illustrated in \fig{fig:down}: the resultant rope does not keep rising after flux feeding (\fig{fig:down}(b)-\ref{fig:down}(c)), but falls back to the photosphere (\fig{fig:down}(d)-\ref{fig:down}(e)), resulting in an equilibrium state in which the rope sticks to the photosphere (\fig{fig:down}(f)). Therefore, if the injected axial flux is not large enough so that the critical axial flux of the major rope is not reached, not only is the flux feeding process unable to cause the major rope to erupt, but also the major rope system is transformed from an HFT configuration to a stable BPS configuration. This result might also suggest a possible scenario for confined flares in HFT configurations.
\par
As demonstrated in \fig{fig:erupt}, obvious magnetic reconnection occurs in the vertical current sheet below the rising major flux rope. Here we simulate a specific case to distinguish between the role that reconnection plays and that upward catastrophe plays in initiating the flux rope eruption: $C_E$ is also 2.23, the same as the eruptive case shown in \fig{fig:erupt}, but the magnetic reconnection in the current sheet below the rope is prohibited. To achieve this, similar simulating procedures as, for example, \cite{Zhang2016a}, are used here: after $t=80\tau_A$, first, the resistivity is adjusted to zero, and second, reassign the flux function $\psi$ along the entire vertical current sheet (if it exists) below the major rope with the initial value $\psi_c=\psi_i(x=0,~y=0)$ at each time step, so that $\psi$ along the current sheet is invariant. With these procedures, both numerical and physical magnetic reconnections below the major rope are prohibited.
\begin{figure*}
\includegraphics[width=\hsize]{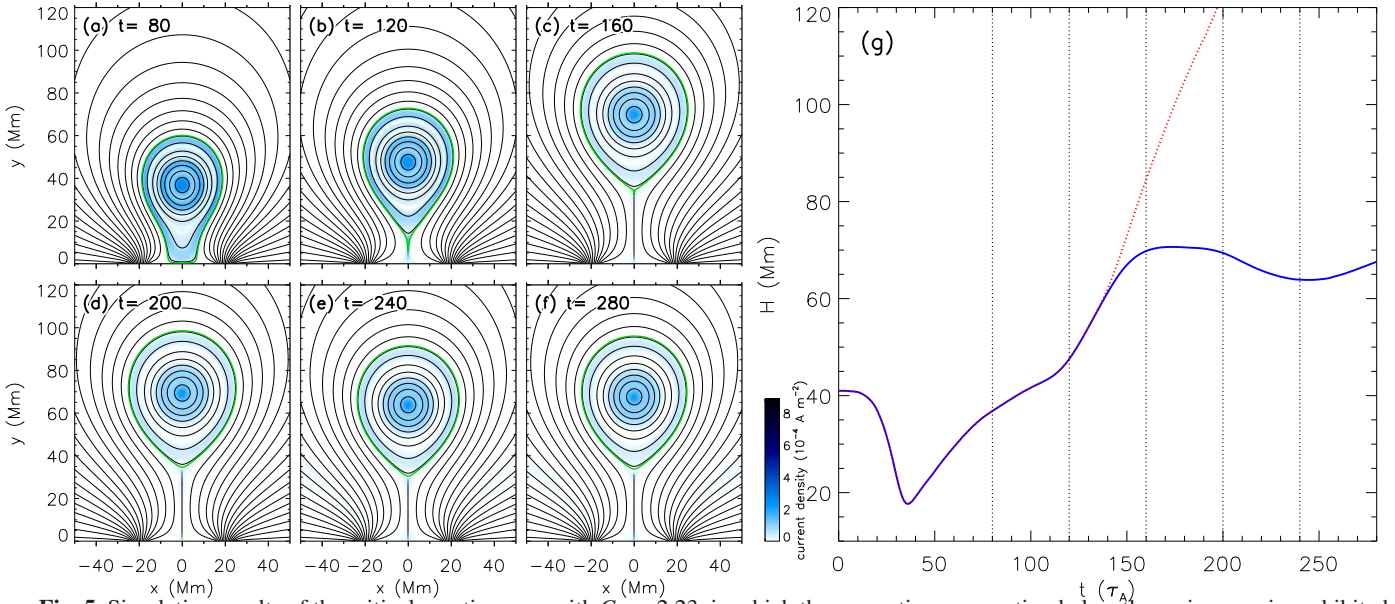}
\caption{Simulation results of the critical eruptive case with $C_E=2.23$, in which the magnetic reconnection below the major rope is prohibited. The meanings of the symbols are the same as those in \fig{fig:erupt}. The red dotted line in panel(g) is the corresponding eruptive profile with reconnection, which is the same as that in \fig{fig:erupt}(i).}\label{fig:without}
\end{figure*}
The simulation results for this specific case are illustrated in \fig{fig:without}. After flux feeding, the upward catastrophe occurs and the resultant rope rises for a while (\fig{fig:without}(a)-\ref{fig:without}(c)), but then stops rising, and eventually, the rope does not erupt but levitates in the corona (\fig{fig:without}(d)-\ref{fig:without}(f)). Without magnetic reconnection, the closed background arcades above the major rope are always rooted at the photosphere. These closed arcades constrain the major rope, so that prevent the rope from further rising. For comparison, in the case with reconnection demonstrated in \fig{fig:erupt}, background arcades above the rope reconnect at the vertical current sheet below the rope, so that the ``tether" constraining the rising major rope is removed. As suggested by \cite{Green2018}, different kinds of eruptive mechanisms can be classified into ``trigger" and ``driver" according to their predominant role in the eruption process. In our simulation results, the flux feeding process causes the upward catastrophe, which triggers the flux rope eruption. Magnetic reconnection as well as the further evolution of the upward catastrophe act as the drivers of the flux rope eruption.

\section{Discussion and conclusion}
\label{sec:dc}
In this paper, we investigate the influence of flux feeding on coronal magnetic flux ropes with an HFT configuration, especially how flux feeding causes the flux rope to erupt. During the flux feeding process, the small emerging flux rope as well as the arcades below the major rope with the HFT configuration reconnect with the pre-existing major flux rope, so that axial magnetic flux is injected into the major rope, whereas the poloidal magnetic flux of the major rope is reduced. Flux feeding is able to cause the major rope to erupt, provided that the amount of the axial flux inject by flux feeding is large enough so that the total axial flux of the major rope reaches its critical value, which is similar as that in the BPS cases investigated in \p. During the early period of the flux feeding process, the major rope gradually descends due to the magnetic cancellation, demonstrating a downward movement of the flux rope right before its eruption, which suggests an alternative theoretical scenario for coronal implosions. Moreover, our simulation reproduce the photospheric magnetic response to the eruption of the coronal flux rope: the horizontal component of the photospheric magnetic field increases around the PIL after the eruption, as is often observed in the wake of major solar eruptions.
\par
Although the onset of the eruption in both the BPS and the HFT cases is dominated by the upward catastrophe associated with the critical axial flux, the influence of flux feeding on the flux ropes with an HFT configuration is different from those with a BPS configuration. With BPS, the major flux rope sticks to the photosphere, so that the small rope directly interacts with the major rope from the very beginning of the flux feeding process, and the major rope only reconnects with the small emerging rope during the flux feeding process; the non-eruptive resultant rope after flux feeding is still with a BPS configuration. For HFT cases, there exisits additional interaction between the major rope and the arcades below it in the initial state. Consequently, not only the poloidal flux of the major rope is reduced due to the reconnection with these arcades, but the HFT configuration collapses after flux feeding. The flux feeding in HFT configurations results in a dramatic change of topology, with the flux rope either erupting or falling back to a BPS configuration. Moreover, previous theoretical studies found that a flux rope could erupt if either its axial or poloidal magnetic flux exceeds the corresponding critical value. For BPS cases, each episode of flux feeding injects axial flux into the flux rope while maintaining its poloidal flux, therefore pushing the rope system one step closer toward the eruption, no matter how small the amount of the injected axial flux is. For HFT cases, however, if the injected axial flux is too little, the reduction in poloidal flux could push the rope system even farther away from the onset of the eruption, and, what's more, the rope system may also evolve from a metastable HFT configuration to a stable BPS configuration after flux feeding. Therefore, the flux feeding process in HFT configurations is not always in favor of the eruption of the major rope; it could even stabilize the rope system.
\par
As introduced in \sect{sec:analysis}, flux feeding cases with different $C_E$ are simulated in this paper. 
The emerging velocity of the small rope, however, is unchanged in different cases. Since the emerging velocity might also have some influence on the evolution of the resultant rope, we will try to simulate flux feeding cases with various emerging velocities in our future work, so as to investigate the role that the kinetic properties of flux feeding processes plays in initiating coronal flux rope eruptions.

\begin{acknowledgements}
We appreciate prof. Jie Zhang for his insightful suggestions and comments. This research is supported by the National Natural Science Foundation of China (NSFC 41804161, 41774178, 41761134088, 41774150, 41842037 and 41574165), the Strategic Priority Program of CAS (XDB41000000 and XDA15017300), and the fundamental research funds for the central universities. We acknowledge for the support from National Space Science Data Center, National Science \verb"&" Technology Infrastructure of China (www.nssdc.ac.cn). 
\end{acknowledgements}


\begin{appendix}
\section{Basic equations and initial preparations}
\label{ape:equations}
Combining the 2.5-Dimensional form of the magnetic field (Eq. \ref{equ:mf}), we could write the MHD equations in the non-dimensional form as:
\begin{align}
&\frac{\partial\rho}{\partial t}+\triangledown\cdot(\rho\textbf{\emph{v}})=0,\label{equ:cal-st}\\
\nonumber &\frac{\partial\textbf{\emph{v}}}{\partial t}+\frac{2}{\rho\beta_0}(\vartriangle\psi\triangledown\psi+B_z\triangledown B_z+\triangledown\psi\times\triangledown B_z)+\textbf{\emph{v}}\cdot\triangledown\textbf{\emph{v}}\\ 
&~~~+\triangledown T +\frac{T}{\rho}\triangledown\rho+g\hat{\textbf{\emph{y}}}=0,\\
&\frac{\partial\psi}{\partial t}+\textbf{\emph{v}}\cdot\triangledown\psi-\frac{2\eta}{\beta_0}\vartriangle\psi=0,\\
&\frac{\partial B_z}{\partial t}+\triangledown\cdot(B_z\textbf{\emph{v}})+(\triangledown\psi\times\triangledown v_z)\cdot\hat{\textbf{\emph{z}}}-\frac{2\eta}{\beta_0}\vartriangle B_z=0,\\
\nonumber &\frac{\partial T}{\partial t}-\frac{4\eta(\gamma-1)}{\rho R\beta_0^2}\left[(\vartriangle\psi)^2+|\triangledown\times(B_z\hat{\textbf{\emph{z}}})|^2 \right]\\
&~~~+\textbf{\emph{v}}\cdot\triangledown T +(\gamma-1)T\triangledown\cdot\textbf{\emph{v}}=0,\label{equ:cal-en}
\end{align}
where
\begin{align}
\vartriangle\psi=\frac{\partial^2\psi}{\partial x^2}+\frac{\partial^2\psi}{\partial y^2},~~\vartriangle B_z=\frac{\partial^2 B_z}{\partial x^2}+\frac{\partial^2 B_z}{\partial y^2}.
\end{align}
Here $\rho$ and $T$ denote the density and the temperature; $v_x, v_y, v_z$ represent the $x$, $y$, and $z-$component of the velocity, respectively; $\gamma=5/3$ is the polytropic index in our simulation; $g$ is the normalized gravity; $\eta$ is the resistivity. Here $\beta_0=2\mu_0\rho_0RT_0L_0^2/\psi_0^2=0.1$ is the characteristic ratio of the gas pressure to the magnetic pressure, where $\rho_0=3.34\times10^{-13}\mathrm{~kg~m^{-3}}$, $T_0=10^6\mathrm{~K}$, $L_0=10^7\mathrm{~m}$, and $\psi_0=3.73\times10^3\mathrm{~Wb~m^{-1}}$ are the characteristic values of density, temperature, length and magnetic flux function, respectively, which are also the calculating units in the simulation. For the other quantities, the corresponding characteristic values are $v_0=128.57$ km s$^{-1}$, $t_0=77.8$ s, $B_0=3.37\times10^{-4}$ T, $g_0=1.65\times10^3$ m s$^{-2}$. The numerical domain in our simulation is $0<x<200$ Mm, $0<y<300$ Mm, which is discretized into 400$\times$600 uniform meshes. At the left side of the domain ($x=0$), symmetric boundary condition is used. The radiation and the heat conduction in the energy equation are neglected. 
\begin{figure*}
\includegraphics[width=1.5\hsize]{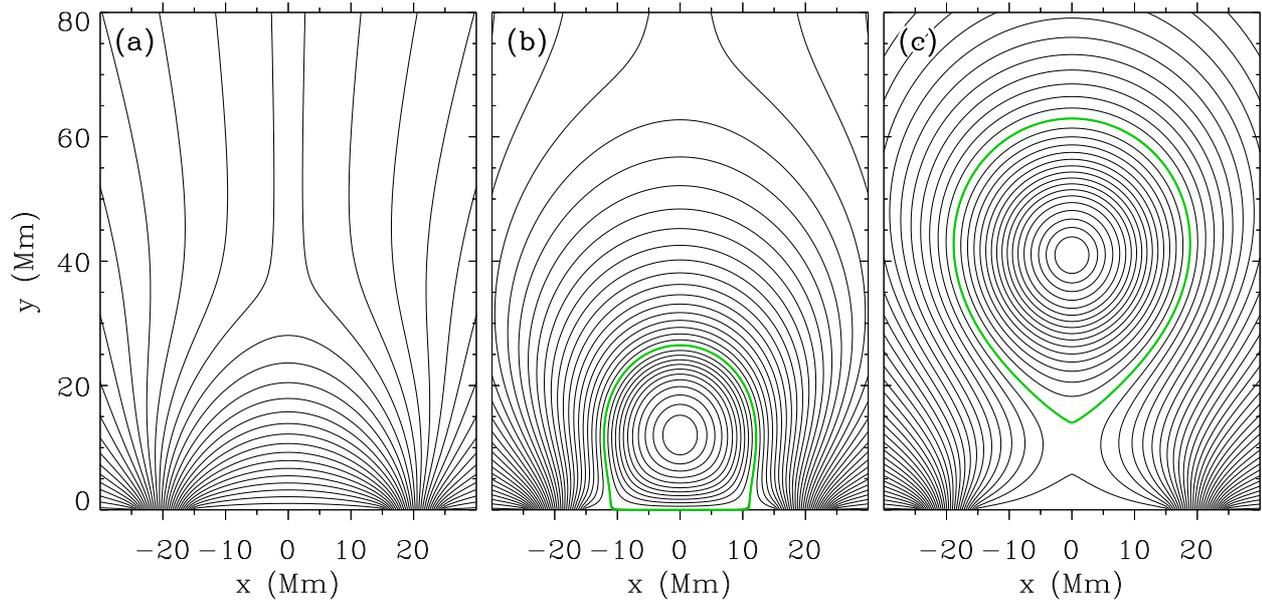}
\caption{Panel (a) is the magnetic configuration of the background field. Panel (b) is an interim state to construct the initial state. Panel (c) is the obtained flux rope system with an HFT configuration, which is the initial state of our simulation.}\label{fig:construct}
\end{figure*}
\par
In order to simulate flux feeding processes in HFT cases, we must first construct a typical coronal flux rope system with an HFT configuration. Here we select a partially open bipolar field as the background field, in which a negative and a positive surface magnetic charge located at the photosphere within $-b<x<-a$ and $a<x<b$, respectively. The background field could then be obtained by the complex variable method \citep[e.g.,][]{Hu1995,Zhang2017a}.
The background magnetic field can be cast in the complex variable form
\begin{align}
f(\omega)\equiv B_x-iB_y=\frac{(\omega+iy_N)^{1/2}(\omega-iy_N)^{1/2}}{F(a,b,y_N)}\mathrm{ln}\left( \frac{\omega^2-a^2}{\omega^2-b^2}\right),
\end{align}
where $\omega=x+iy$, and
\begin{align}
\nonumber &F(a,b,y_N)=\frac{1}{b-a}\int_a^b(x^2+y_N^2)^{1/2}dx=\frac{1}{2(b-a)}\times\\ &\left[b(b^2+y_N^2)^{1/2}-a(a^2+y_N^2)^{1/2}+y_N^2\mathrm{ln}\left(\frac{b+(b^2+y_N^2)^{1/2}}{a+(a^2+y_N^2)^{1/2}} \right)\right].
\end{align}
Here $a=15$ Mm, $b=25$ Mm, and ($y=y_N=34.5$ Mm, $x=0$) is the position of the neutral point of the partially open bipolar field. There is a neutral current sheet located at $(x=0,~y\geq y_N)$. The magnetic flux function could then be calculated by:
\begin{align}
\psi(x,y)=\mathrm{Im}\left\lbrace\int f(\omega)d\omega \right\rbrace,\label{equ:integral}
\end{align}
and the flux function at the lower base is
\begin{equation}
 \psi_i(x,0) = \left\{
              \begin{array}{ll}
              {\psi_c}, &{|x|<a}\\
              {\psi_c F(|x|,b,y_N)/F(a,b,y_N)}, &{a\leqslant|x|\leqslant b}\\
              {0}, &{|x|>b}
              \end{array}  
         \right.\label{equ:fluxb}
\end{equation}
where $\psi_c=\pi\psi_0$; the flux function at the neutral point $y=y_N$ can be calculated as:
\begin{align}
\psi_N=\frac{\pi(b^2-a^2)}{2F(a,b,y_N)}.\label{equ:fluxc}
\end{align}
Letting $B_z$ equals 0 in the background field, the configuration of the background field could then be obtained, as shown in \fig{fig:construct}(a). The initial corona is isothermal and static:
\begin{align}
T_c\equiv T(0,x,y)=1\times10^6 ~\mathrm{K},\ \  \rho_c\equiv\rho(0,x,y)=\rho_0\mathrm{e}^{-gy}.\label{equ:rhot}
\end{align}
Except during the emergence of the small rope, the lower boundary is fixed: the flux function $\psi$ is fixed at $\psi_i$ given by Eq. \ref{equ:fluxb}; $B_z$ and the velocity is always 0; the density and the temperature are fixed at their initial values. Increment equivalent extrapolation is used at the right and top boundaries.
\par
With the initial condition and the background configuration, equations (\ref{equ:cal-st}) to (\ref{equ:cal-en}) are solved by the multi-step implicit scheme. First, we let a flux rope emerge from the lower base of the background field, as shown in \fig{fig:construct}(b). By multiplying $B_z$ within the rope by a factor larger than 1, we increase the axial magnetic flux of the flux rope. The rope then rises, and magnetic reconnection occurs within the veritical current sheet formed below the rope, resulting in closed arcades below the rope. After that, adjust the axial flux of the rope again, so that the rope stops rising. Eventually, let the rope system relax to a equilibrium state with an HFT configuration, as shown in \fig{fig:construct}(c). This is the initial state of our simulation. It is noteworthy that the radius of the flux rope in our simulation is finite, so that the thin-rope approximation is not satisfied. Under this circumstance, the initial state could only be obtained by numerical procedures.
\end{appendix}

\end{document}